\documentstyle[eqsecnum,preprint,aps,prd]{revtex}
\preprint{ITP-96-66, UCSBTH-96-20}
\date{\today}
\draft
\tighten
\begin{document}
\def\sqr#1#2{{\vcenter{\hrule height.3pt
      \hbox{\vrule width.3pt height#2pt  \kern#1pt
	 \vrule width.3pt}  \hrule height.3pt}}}
\def\square{\mathchoice{\sqr67\,}{\sqr67\,}\sqr{3}{3.5}\sqr{3}{3.5}}
\def\today{\ifcase\month\or
  January\or February\or March\or April\or May\or June\or July\or
  August\or September\or October\or November\or December\fi
  \space\number\day, \number\year}

\def\Bbb{\bf}


\title{Bubbles in Kaluza-Klein theories with space- or time-like internal
dimensions}

\author{A. Chamblin{$^1$} and R. Emparan{$^2$}}
 
\address {\qquad \\ {$^1$}Institute for Theoretical Physics\\
University of California\\
Santa Barbara, California 93106-4030, U.S.A.
\qquad\\{$^2$}Dept. of Physics\\
University of California\\
Santa Barbara, CA 93106, U.S.A.
}

\maketitle

\begin{abstract}

Bubbles are point-like regular solutions of the higher-dimensional
Kaluza-Klein equations 
that appear as naked singularities in four dimensions. 
We analyze all such possible
solutions in 5D Kaluza-Klein theory that are static and 
spherically symmetric.
We show that they can be obtained by taking unusual choices 
of the parameters in the dyonic black hole solutions,
and find that regularity can only be achieved if their electric charge
is zero. However, they can be neutral or possess magnetic charge. We study
some of their properties, both in theories where the internal dimension is
space-like as well as time-like. Since bubbles do not have horizons, 
they have no entropy, nor do they emit any thermal radiation, 
but they are, in general, non-extremal objects. In the two-timing case, it 
is remarkable that non-singular massless monopoles are possible, probably 
signaling a new pathology of these theories. These two-timing monopoles 
connect two
asymptotically flat regions, and matter can flow from one region to the
other. We also present a C-type 
solution that describes neutral bubbles in uniform acceleration, 
and we use it to construct an instanton that mediates the breaking 
of a cosmic string by forming bubbles at its ends. The rate for this process
is also calculated.  Finally, we argue that a similar
solution can be constructed for magnetic bubbles, and that it can be used
to describe a semiclassical instability 
of the two-timing vacuum against
production of massless monopole pairs.

\end{abstract}


\section{Introduction}

The solitonic solutions of the gravitational
field equations have always been a focus of interest, 
much more in recent times after the recognition
of their relevance towards a non-perturbative understanding of string theory
\cite{DUFF}. 
We refer to solitons in a broad sense,
as localized solutions of the classical equations of motion that are
non-singular or possibly have singularities hidden behind a horizon, the 
latter corresponding to black holes, or more generally, black branes.
Certainly, to qualify as particle states one would also like the solutions
to be stable.
However, it has been proven for a 
wide range of solutions
that non-extremal $p$-branes (with $p\geq 1$) are classically
unstable \cite{GREGORY}. Nonetheless, though unstable, many of these 
solutions are interesting on their own.

Most of the solitons that have been considered are of black hole type, 
i.e., objects
with horizons and, correspondingly, an associated entropy
---in some cases, though, the horizon area vanishes in the
extremal limit, and singularities appear. However, there also exists
an interesting class of 
solutions that, even if they look like naked singularities in 
four dimensions, their singularity can be resolved by going to a 
higher dimension.
This resolution of singularities is purely classical, not involving
any quantum smearing of the region surrounding the singularity, and it 
has been shown to smooth the inner singularity inside some black holes and
black branes of interest \cite{GHT}. If applied to resolve naked 
singularities,  the outcome is a completely regular geometry 
for a point-like object, 
though, as we will see,  it is often an unstable one. Throughout the paper we
will be working in the simplest theory in which bubbles can appear, namely
five-dimensional Kaluza-Klein theory. Many of the features described here
should also be present in more general theories with compact dimensions. 

We will describe two ways in which a naked singularity can be regularized when
`blown up' to reveal the higher dimensional
structure of spacetime. In one situation, the internal
space closes up smoothly ---in the microscopic region representing the
point-particle--- avoiding in this way that the singular origin 
could be reached. Geodesic
completeness is preserved by leaving a non-singular `hole' in space. 
In the other case, the pointlike particle turns out to
be a tiny microscopic region bridging two asymptotically flat spacetimes
through ``chronology horizons''.

Solutions of this kind
have been known for some time as ``bubbles'', and this is the name we 
will adopt throughout this paper. An early realization
of the importance of bubbles was found by Witten in \cite{WITTEN}, 
where it was argued that the Kaluza-Klein vacuum can decay by 
spontaneously forming bubbles that exponentially expand after their creation. 
In this paper, however, we will exclusively focus on non-expanding bubbles.
An interesting feature of all bubbles is
that they do not have
horizons, and therefore are zero-temperature objects with zero entropy.

Bubbles appear when the internal isometry along which the dimensional 
reduction is performed contains fixed points. The 
Gross-Perry-Sorkin (GPS) monopole \cite{SORKIN,GROSS} is an example
of a bubble in which the fixed point set consists 
of a single point 
and the bubble has zero size. In general, however, the fixed 
point set will be higher
dimensional, and the bubble will have finite size. 
Several bubble solutions of 5D Kaluza-Klein theory (with spacelike internal 
dimension) were already noticed 
in \cite{SORKIN,GROSS}, and
also discussed in \cite{GIBBONS1,GIBBONS2}. In this paper we aim at 
determining in a complete way all
the possible static spherical bubble solutions of 5D Kaluza-Klein theory, 
clarify
their relationship with the already thoroughly studied black hole 
solutions of the
theory, and analyze bubble pair production and its consequences.
Parallelly, we will study a much less explored issue, that of the
solitonic objects 
in KK theories in which the internal dimension is timelike.

Theories with internal timelike dimensions
have been a subject of  interest for some time, and, 
in a rather modified way, have been recently revived in the context of string 
theory (`F-theory')\cite{VAFA}. However, not much is known about the
classical vacuum solutions of two-timing theories. Recently, a full 
classification of the possible isometry groups of compact 
Lorentzian manifolds has
been achieved \cite{ADAMSTUCK}, and this may be of relevance in determining
the properties of compactifications in such theories. For the moment,
however, we will 
restrict ourselves
to the simplest case of a single internal timelike $S^1$.

The timelike character of an internal direction gives rise to several 
pathological features, the most conspicuous of which may be the fact that 
excitations of the internal
dimensions have negative norm. Experimental lower bounds on possible
violations of unitarity put a limit on the maximum radius of
the internal timelike direction \cite{YNDURAIN}. If one wants to preserve 
unitarity in a higher energy range, these
degrees of freedom must be somehow frozen out or gauged away. An additional
problem is that, as we will see in a moment, the
coupling constant coming from the compactification modulus is imaginary. 
Oppositely charged objects then repel, and,
by an old argument due to Dyson,
one would expect the vacuum to be unstable against pair production
of charged particles\cite{DYSON}. An instability of this kind was sought in 
\cite{RASHEED}, where it was noticed that the 
two-timing theory contains massless black
holes, and solutions describing a pair of them accelerating apart can be
found. However, due to peculiar thermal properties of these black holes,
an instanton describing their pair creation could not be constructed. One of 
the aims of this paper is to show that this obstruction can
be surmounted if instead of massless black holes one considers {\it massless
monopole bubbles}, to be constructed below.

Our analysis of the bubble solutions in theories with internal time will
be greatly simplified by noticing their relation with solutions with internal
space: 
given a solution of the 5D Kaluza-Klein equations with spacelike internal
dimension,
\begin{equation}
ds^2=e^{-4\phi/\sqrt{3}} (dx^5 +2 A_\mu dx^\mu)^2 
+e^{2\phi/\sqrt{3}}g_{\mu\nu}dx^\mu dx^\nu
\end{equation}
where $\phi$ is the scalar modulus (`dilaton'), 
$A_\mu$ the Maxwell four-potential, and 
$g_{\mu\nu}$ the effective four dimensional metric,($\mu,\nu=0,\dots,3$), 
a solution of the two-timing
theory can be readily obtained by Wick-rotating
\begin{equation}
x^5\rightarrow ix^5,
\end{equation}
and, accordingly, in order to keep the solution real,
\begin{equation}
A_\mu\rightarrow iA_\mu.
\end{equation}
The latter amounts to making  $e^2$, the square of the electric charge,
take on negative values, i.e., the 
coupling constant has been continued to an imaginary value and objects with 
the same charge attract each other.
This method will allow us to easily generate bubble solutions of two-timing 
theories. However, even if formally similar, the two-timing bubbles will
differ considerably in their properties from ordinary bubbles.

\section{Bubbles in 5D Kaluza-Klein with internal spacelike direction}
\label{sec:KKmnpls}

We want to find all the topologically non-trivial regular solutions of
5D KK theory which appear as static, spherically symmetric, points acting 
as sources of the Maxwell field. 
Complete analyses of the static, spherically symmetric solutions of 
5D KK theory
have been given in different forms, see e.g. \cite{DOBIASCH,CVETIC}, and 
also \cite{WILTSHIRE}, but the focus has been generally on the black hole
solutions. However, we will find that bubble solutions 
correspond to unconventional
choices of the parameters in black hole solutions with both electric and
magnetic charge. It is convenient to illustrate this first in the 
simplest example where the magnetic charges are zero.

It has been known for some time that, by 
taking the Schwarzschild solution and boosting it in the fifth direction,
the entire family of electrically charged (magnetically neutral)
KK black holes is generated. The five dimensional metric in this case is
\begin{eqnarray}
\label{eqn:boosted}
ds^2 &=& \left( 1+{2m\sinh^2\delta_Q\over r}\right) (dx^5 +2 A_t dt)^2
-{1-2m/r\over 1+{2m\sinh^2\delta_Q/ r}}dt^2 +{dr^2\over 1-{2m\over r}} 
+r^2d\Omega^2_2,\\
A_t &=& {m\sinh 2\delta_Q\over 2r( 1+{2m\sinh^2\delta_Q/ r})}.\nonumber
\end{eqnarray}
Here the electric charge $Q$ is a function of the boost parameter 
$\delta_Q$:
\begin{equation}
Q={m\over 2}\sinh 2\delta_Q= m{v\over 1-v^2},
\end{equation}
where $v$ is the boost velocity. By taking $0\leq v <1$ we obtain, upon
reduction, the spectrum of purely electric black holes. The limit 
$v\rightarrow 1$ (which requires $m\rightarrow 0$, keeping the charge
and mass finite) describes an extremal singular solution. 

What is less well known is what happens when we take $v>1$. Amusingly, 
even if the boost parameter $\delta_Q$
in these solutions is complex, with imaginary part $i\pi/2$, we still find 
real solutions.
The four dimensional reduced metric does not contain a horizon any more,
but, rather, a naked singularity. However, we must analyze if this 
singularity is also present in the
full five dimensional geometry. 

For all the values $1<v<\infty$ we have an electrically charged solution for
which the internal circles close at 
\begin{equation}
r=r_B \equiv 2 m{v^2\over v^2 -1}.
\end{equation}
This closing of the internal space is the characteristic feature of bubbles
with internal spacelike directions.
However, for any finite $v>1$ the curvature is singular at $r_B$. 
This can be easily seen from the
fact that, even if the length of the internal circles goes to zero at
that point, their proper radius does not. Additionally, the $g_{tt}$ term diverges
badly at the bubble.

The singularity disappears only for $v\rightarrow \infty$. Of course, 
this is equivalent to interchanging the roles of the
internal and time coordinates in the Schwarzschild solution:
\begin{equation}
\label{eqn:hole}
ds^2=-dt^2 +\left( 1-{2 m\over r}\right) (dx^5)^2 + {dr^2
\over 1-2m/ r} + r^2 d\Omega_2^2.
\end{equation}
This is simply the product of the Euclidean Schwarzschild solution with a
trivial time direction, and is a completely nonsingular geometry.
It describes a neutral ``bubble'': points $r< 2m$ are excised from 
spacetime, but this is done in a nonsingular way: 
if the periodicity of $x^5$ is chosen to be 
$8\pi m$, the internal space smoothly closes at $r=2m$. 
In the reduced description, it corresponds to a pointlike naked singularity,
with ADM mass equal to $m/2$. 

Notice that the Schwarzschild solution, and all the electrically charged
black holes can in turn be obtained by boosting
the neutral bubble with velocity $v>1$. In a sense, the Schwarzschild solution and
the neutral bubble are at opposite endpoints of the electrically charged
spectrum. This will also be true when we add magnetic charge.

The general static, spherically symmetric
solution with both electric and magnetic charges can be found in
different parametrizations in \cite{DOBIASCH,WILTSHIRE,CVETIC}. We find
it convenient to use the one in \cite{CVETIC}, which is given in terms
of the two boost 
parameters applied to the Schwarzschild solution with mass $m$, 
since it is more closely related to the analysis 
above. For completeness, we discuss in Appendix \ref{apa} how the bubbles
appear when the metric coefficients are expressed directly in terms of 
the physical parameters $M,P,Q$ 
and the scalar charge as in \cite{WILTSHIRE}.

The five dimensional geometry
of the solutions is \cite{CVETIC}
\begin{equation}
\label{eqn:gensol}
ds_{(5)}^2 = {X\over Y}(dx^5 + 2A_\mu dx^\mu)^2 -{f\over X}dt^2 +
{Y\over f} dr^2 +Y d\Omega_2^2,
\end{equation}
\begin{equation}
\label{eqn:fourpot}
A_t= Q{r+2m\over X},\qquad A_\varphi = P \cos\theta,
\end{equation}
where
\begin{eqnarray}
\label{eqn:XYf}
X &=& (r +2m)(r+ 2m \cosh^2\delta_Q),\nonumber\\
Y &=& r^2 +2m[(2-\cosh^2\delta_Q) \cosh^2\delta_P +\cosh^2\delta_Q]r
+4m^2\cosh^2\delta_P,\\
f &=& r(r+2m),\nonumber
\end{eqnarray}
and the electric and magnetic charges and ADM mass are
\footnote{Our conventions differ slightly from \cite{CVETIC}: 
with our definitions, the Maxwell field is one half of theirs, and our (ADM)
mass is one fourth of their parameter $M$. Also, what we call $\delta_P$ 
is their $\delta_1$, and our $m$ is their $\beta$.}
\begin{eqnarray}
Q &=& m \sinh \delta_Q\cosh\delta_Q\cosh\delta_P,\nonumber\\
P &=& m (2-\cosh^2\delta_Q)\sinh \delta_P\cosh\delta_P,\\
M &=& {m\over 2}(\cosh^2\delta_Q +\cosh^2\delta_P 
-\sinh^2\delta_Q\sinh^2\delta_P).
\end{eqnarray}
Here $m$ corresponds to the mass parameter of the Schwarzschild solution
used to generate the whole family. Extremal solutions correspond to limits
where $m\rightarrow 0$, while sending a boost angle to infinity to keep
the mass and some of the charges fixed\footnote{There is a family of
neutral solutions \cite{GROSS}, of which the Schwarzschild 
black hole and the neutral bubble are particular cases, that is not
contained in general in \cite{CVETIC}. However, neither these nor their 
boosted counterparts yield new regular solutions.}. To avoid Taub-NUT terms
in the reduced metric we must keep $m\sinh 2\delta_Q\tanh\delta_P =0$.

The horizons correspond to the zeroes of $f$ that are not simultaneously 
zeroes of 
$X,Y$. For $\cosh^2\delta_Q\geq 1$ there is always a horizon at $r=0$. 
These
solutions are black holes, and therefore we will 
concentrate on solutions with electric boost 
velocity greater than $1$, i.e.,
$\cosh^2\delta_Q\leq 0$. Bubbles will appear when $f$ and $X$ have a common
zero, which must also be bigger than the roots of $Y$ since these, 
in general, are 
singularities. However, just as in the magnetically neutral case, and for the
same reasons, the
metric will be singular if there
is a nonvanishing electric term. Thus we shall set
$\cosh\delta_Q=0$. 
Notice that the neutral solution obtained by setting the electric and 
magnetic charge to zero by making $\cosh\delta_Q=\cosh\delta_P=0$ is singular.
Therefore we consider $\cosh^2\delta_P\neq 0$.

Remarkably, when $\cosh\delta_Q=0$ we have $f=X$
and the 5-metric is the
direct product of a `trivial' time with a Euclidean 
four-geometry\footnote{This is more 
apparent if we write the metric (\ref{eqn:gensol}) in the reduced form of 
\cite{DOBIASCH} on which the analysis of \cite{CVETIC} is based.}. 
The latter
is, in fact, the Euclidean Taub-NUT metric. 
It can be written in a more familiar way by 
shifting $r+2m\cosh^2 
\delta_P\rightarrow r$ and then identifying the
magnetic charge $P$ as the Euclidean nut parameter:
\begin{equation}
\label{eqn:5DTN}
ds_{(5)}^2=-dt^2 +U(r)(dx^5+ 2P\cos\theta d\varphi)^2 +{dr^2\over U(r)}
+(r^2-P^2)d\Omega_2^2,
\end{equation}
\begin{equation}
U(r)={r^2-4M r+P^2\over r^2-P^2}.
\end{equation}  
(we stress that the parameter $M$ here is the mass of the monopole, and not 
the usual `mass parameter' of the four dimensional
Taub-NUT metric, which is in fact $2M$).
We still have to impose further regularity conditions on this solution, 
but for the moment note that this covers all the static, 
asymptotically flat, spherically symmetric, regular 
solutions. They represent localized
lumps of energy which are sources of a monopolar magnetic field.

The factorized form of the geometry (\ref{eqn:5DTN}) greatly simplifies
the analysis of the regularity of the solutions.
There are only two completely non-singular instantons constructed out of
the Euclidean Taub-NUT solution: the self-dual Taub-NUT instanton, and the 
Taub-bolt instanton. The Eguchi-Hanson metric can be obtained as a limiting 
case of Taub-NUT, but, in the KK context, it does not yield an 
asymptotically flat reduced geometry.

The self-dual Taub-NUT solution corresponds to a minimal value for the mass,
$M_N=P/2$, and yields just the GPS monopole, which is
located at the nut fixed
point of $\partial_5$. If we allow the mass of the monopole to be greater 
than the minimal value
$M_N$, then, as shown by Page \cite{PAGE},
there will be a two dimensional fixed point set of $\partial_5$,
(a spherical bolt) at the
biggest root  of $U(r)$, $r_+=2M +\sqrt{4M^2-P^2}$. 
A conical singularity at the bolt
can be avoided by
choosing $M=5 P/8$. The metric (\ref{eqn:5DTN}) with
\begin{equation}
 \label{eqn:Taub-bolt}
U_B(r) = \frac{(r-r_+)(r-r_-)}{r^2-P^2},\qquad
r_+ = 2P$, $r_- = \frac{P}{2},
\end{equation}
is referred to as
the `Taub-bolt' solution. It is not
self-dual, and is
topologically ${\Bbb C}{\Bbb P}^2 - \{ 0 \}$.  The point which has been
removed from
${\Bbb C}{\Bbb P}^2$ is `infinity' ($r \rightarrow \infty$).

For all other values of $M>P/2$, $M\neq 5P/8$, we obtain a conical 
singularity in the internal
direction. This is a very mild singularity, and we could change the
periodicity of $\tau$ to be $\Delta\tau=4\pi r_+$
so as to avoid the conical singularity at the bolt, but
at the same time we should accordingly change the periodicity 
of $\varphi$ to
$\Delta\varphi= \pi r_+/P$ to avoid the Dirac string singularities.
In turn, this would result in a conical defect
\begin{equation}
\delta=2\pi\left( 1 -{r_+\over 2P}\right)
\end{equation} 
threading the monopole along the axes $\theta=0,\pi$.
For $P/2<M<5P/8$ the defect is a string, whereas  
for $M>5P/8$ it is a strut. Therefore, the former case can represent, 
if not new particle-like solutions, configurations 
of bolt monopoles threaded with cosmic string vortices, reminiscent of
similar black hole/cosmic string configurations considered in
\cite{ACHUCARRO}.

It is amusing that the GPS monopole 
appears as an extremal limiting case
of two kinds of objects: the magnetic black holes and the magnetic bubbles.
The black holes can be regarded as thermally excited states of the monopole, 
to which they decay by emitting Hawking radiation. 
The bolt monopole is likely to
correspond to another type of excited state, but since it has no horizon it
does not emit Hawking radiation. However, it is known to be classically
unstable, since the spectrum of fluctuations around the Euclidean Taub-bolt 
metric contains a negative mode\cite{YOUNG}, 
as does the Euclidean Schwarzschild solution. It has been
speculated \cite{GIBBONS2} that bolt monopoles should evolve to form a 
black hole, which would then decay to the GPS monopole. 
In any case, the most important restriction on the possible
relevance of bolt monopoles comes from the fact that fermions rule them out,
since the topology  ${\Bbb C} {\Bbb P}^2-\{0\}$
does not admit any spin structure.

An interesting outcome of the analysis in this section is the 
place we have found for the bubbles among the 
general spectrum of dyonic solutions. This will allow us to extend to bubbles
the study of some processes involving black holes. Before that, we will
analyze some striking features of the bubble solutions when the signature
of the internal dimension is reversed.

\section{Two-timing monopoles}

As we saw in the introduction, given a solution to ordinary KK theory
we can easily obtain another solution 
corresponding to a two-timing 
theory. However, in general, their properties, such as the nature of their 
singularities, are markedly different. 

It is straightforward to see that 
solutions with electric charge have, again, curvature
singularities. On the other hand, the magnetic bubbles we have just 
described turn into products of the
{\it Lorentzian} Taub-NUT solution with a trivial time,
\begin{equation}
\label{eqn:2tTN}
ds_{(5)}^2=-dt^2 -U(r)(dx^5+ 2P\cos\theta d\varphi)^2 +{dr^2\over U(r)}
+(r^2+P^2)d\Omega_2^2,
\end{equation}
\begin{equation}
U(r)={r^2-4M r-P^2\over r^2+P^2},
\end{equation}  
which is a solution with full five dimensional signature $(- - + + +)$.
After reduction, this describes an object with (imaginary) magnetic charge 
$P$ and mass $M$.

The metric (\ref{eqn:2tTN}) has singularities at 
\begin{equation}
r = {r_\pm}=2M \pm \sqrt{P^2 + 4M^2}.
\end{equation} 
Notice that $r_+>0$ and $r_-<0$. There is now an important difference with the
one-timing solutions. 
As before, the length of the internal space generated by $x^5$ goes to zero 
at $r_\pm$, but, as pointed out in
\cite{HAWKING}, spacetime is easily extended through these surfaces, 
in a way analogous
to the extension through a Lorentzian horizon. We summarize 
the causal structure of the 
maximal extension:
\begin{enumerate}

\item In the region $r_- \,{\leq}\, r \,{\leq}\, r_+$, the coordinate $r$
is
timelike and there are no closed timelike curves (CTCs).

\item In the regions $r \,{\leq}\, r_-$ and $r \,{\geq}\, r_+$, the
coordinate
$x^5$ is timelike and there are (microscopic) CTCs. 

\end{enumerate}

Thus, the surfaces $r = {r_\pm}$
are in
fact chronology horizons. At $r = {r_\pm}$, $U(r) = 0$ and thus the `length' of
the `internal space' generated by $x^5$ goes to zero there. 
Notice that the microscopic bubble region (1) connects two asymptotically 
flat regions. 

Thus, we see that our two-timing Kaluza-Klein monopole can also be regarded 
as a `bubble' of
Kleinian signature, which exists for all external time, and which is bounded
from our
part of the universe (where things are effectively Lorentzian) by a horizon
at $r
= r_+$.
Observers `far away' from the source
($|r| >>
r_\pm$) will see a magnetic monopole. As we approach 
$r_+$ (from
$r >
r_+$), the size of the internal space shrinks to zero at $r=r_+$. 
As we move into the
region $r_-<r < r_+$ 
we see that the internal space is now spacelike! The
`time'
which used to live in the internal space is `uncompactified' in the region 
$r_-<r < r_+$,
and so this region (to a five-dimensional observer, the region is too 
small to be 
detected by a four-dimensional observer) would look like a `bubble'
of
Kleinian signature, i.e., in the ill-defined effective 
`four-dimensional theory', the metric
seems
to change signature from $(- - + +)$ to $(- + + +)$ as we move from the region
$r_-<r <r_+$ to the region $r > r_+$. 

The magnetic charge of these solutions is, as we saw in the introduction, 
imaginary,
and like-charged monopoles attract, rather than repel each other.
No non-singular neutral two-timing bubble exists: 
the magnetic charge $P$ is essential in allowing the extension of
the metric through the interior of the bubble by preventing the formation
of a curvature singularity inside it. 
Thus, the neutral solution with $P=0$ contains 
the usual Schwarzschild singularity at $r=0$. There are more respects in which 
the spectrum of two-timing bubbles differs from ordinary one-timing bubbles. 
For the latter, there is a state with minimal 
non-vanishing mass. This is the GPS monopole, which is extremal and
saturates a Bogomolnyi bound. No extremal state exists for two-timing 
monopoles,
and two-timing multi-monopole static configurations necessarily contain 
conical singularities reflecting the presence of interactions between them.

There are more differences between both classes of objects. 
The background geometry of two-timing monopoles admits fermion structures. 
More strikingly, in the two-timing case 
the value of $M$ is completely unrestricted, 
and we can perfectly well set $M=0$. 
Thus we find a regular {\it massless monopole}.
This case is also specially interesting since the
two different regions $r > P$ and $r < -P$ connected by the Kleinian bubble 
are isometric under $r\rightarrow -r$. 

In the massive case the isometry is broken. We can also have
negative mass objects. This is surprising, since negative mass 
usually implies
naked singularities. These are certainly present in the four dimensional
description 
of these monopoles, but the full solutions are regular. 
The significance of
these negative mass solitons is unclear, but the apparent absence 
of a lower energy bound is probably one more indication of the pathological
behavior of these theories. 

The classical stability of these 
two-timing solutions cannot be inferred from that of their one-timing
counterparts, and this remains an interesting problem. The fact that 
like-charges attract each other suggests that these monopoles would 
tend to a minimal size. 
Let us use this admittedly
non-rigorous criterion to analyze which monopole configuration
is likely to be more stable. 
The monopoles have two ``mouths'', one at $r=r_+$, the other
at $r=r_-$. The total area of both mouths is 
$A \propto (r_+^2 +r_-^2 + 2P^2)= 16 M^2 + 4P^2$, so that, for fixed $P$, the
minimal value is attained when $M=0$. This would suggest that the massless
monopole might be, at least, more stable than the massive states, with
masses of either sign. Whether these solutions can evolve to a black
hole state is not clear, given the strange properties of black holes
in these theories \cite{RASHEED}.

It is of some interest to calculate geodesic motion in the 
background of
the two-timing monopole. To this end, note that from the metric 
(\ref{eqn:2tTN}) we recover, via
the action principle, the following Lagrangian:
\begin{equation}
L = - {\dot t}^{2} - U(r) \left({\dot x_{5}} \,+\, 2P{\cos}{\theta}
{\dot \varphi}\right)^{2}
\,+\, \frac{1}{U(r)} \,{\dot r}^{2} \,+\, \left(r^{2} \,+\, P^{2}\right) 
\left({\dot \theta}^{2} \,+\, {\sin}^{2}{\theta}{\dot \varphi}^{2}\right)
\end{equation}
{where `{\bf $\cdot$}' denotes differentiation relative to any 
affine parameter
$s$. As usual, for photons $L = 0$ whereas for massive particles $L > 0$. 
Now, $L$ is cyclic in $t$, $x_{5}$ and $\varphi$, and so we obtain three
constants of motion:
\[
\left\{ \begin{array}{l}
\varepsilon = {\dot t} = \mbox{`energy'} \\
 \\
\omega = \left(r^{2} \,+\, P^{2}\right) \,{\dot \varphi} = 
\mbox{`angular momentum'} \\
 \\
\delta = U(r) \,{\dot x_{5}} = \mbox{`5-dimensional velocity'}
\end{array} \right.
\]
The 5-dimensional velocity is interpreted by the 4-dimensional observer as
electric charge. 

{}For simplicity, we restrict our attention to the motion of 
test particles in the
${\theta} = \frac{\pi}{2}$ plane, i.e., ${\dot \theta} = 0$ for all times. 
Thus,}
\begin{equation}
L = - {\dot t}^{2} - U(r) \,{\dot x_{5}}^{2} \,+\, 
\frac{1}{U(r)} \,{\dot r}^{2} \,+\, \left(r^{2} \,+\, P^{2}\right) 
\,{\dot \varphi}^{2}
\end{equation}
To be concise, we will only consider the motion of light-like particles, 
`photons'; the timelike case is very
similar. Thus, assuming $L = 0$ we see that the Lagrangian assumes the form
\[
\left( \frac{dr}{d\varphi} \right)^{2} = U(r) \left(r^{2} 
\,+\, P^{2}\right) \left[
\frac{{\varepsilon}^{2}}{{\omega}^{2}} \,+\, 
\frac{{\delta}^{2}}{U(r){\omega}^{2}} - 1
\right]
\]
Radial motion of the photon is possible only if $\left( \frac{dr}{d\varphi}
\right)^{2} > 0$, i.e., if and only if
\[
{\varepsilon}^{2} > {\omega}^{2} - \frac{{\delta}^{2}}{U(r)}
\]

A natural question is whether or not it is possible for a photon to 
escape out to
infinity $(r = + \infty)$ from the monopole `core' (i.e., $r = r_{+}$). 
Of course, the
answer to this question is trivial once one realises that at $r = r_{+}$, 
$U(r_{+}) =
0$! That is, the above inequality becomes
\[
{\varepsilon}^{2} > {\omega}^{2} - \infty
\]
which is trivially satisfied for any values of the energy and angular 
momentum, as long as the `charge' ${\delta}$ is non-vanishing (even when
${\delta}$ is non-vanishing we can just set the angular momentum to zero). Thus,
any photon starting in the core will always make it out to $r = \infty$. 
Likewise,
since any photon beginning in the asymptotic region will trivially satisfy
$\varepsilon > - \infty$, any such particle can fly through the monopole 
core and
emerge in the other asymptotic region.

{}From the point of view of quantum scattering, this presumably means 
that the monopole
is stable to quantum corrections as long as the flux from one 
asymptotic region is
exactly compensated by the flux from the other region. However, 
as we discussed above,
the two asymptotic regions are isometric if and only if $M = 0$. 
This would seem to provide further support to the conjecture
that the $M = 0$ state would be `stable', 
and that massive states would want to
`settle down' to $M = 0$. 

However,  
the massless two-timing
monopole is neither extremal nor does it saturate
any Bogomolnyi-type bound. Therefore we do not expect to find that the 
background admits
$N=1$ supersymmetry.  In other words, there should not exist 
any covariantly constant
spinor field ${\epsilon}$ (by covariantly constant we mean 
constant relative to the Levi-Civita connection on Taub-NUT).  
To see that this is true 
simply recall that the only
Lorentzian manifolds admitting covariantly constant spinor fields are 
the pp-waves; however, all
pp-waves have Type-N curvature, whereas Taub-NUT is of Type-D.
In view of this, the precise 
meaning of our vague
arguments about the stability of the solutions is at present unclear.

\section{Pair creation of neutral bubbles}

\subsection{C-metric for accelerating bubbles}

A number of ways of pair producing black holes have been studied in recent
years. In Sec.~\ref{sec:KKmnpls} we have seen that Kaluza-Klein 
black holes and bubbles form part of the same family of dyonic solutions.
This will allow us to obtain an instanton describing pair creation
of neutral bubbles. They will appear at the ends of a cosmic
string that snaps into two pieces; after that, the two neutral bubbles
accelerate away pulled by the string tension.
Notice that the same argument that allows a topologically stable
string vortex to end in a black hole 
\cite{ACHUCARRO,EARDLEY} can 
be extended to a neutral bubble: the non-trivial topology ${\bf R}
\times S^2\times {\bf R^2}$ of the neutral bubble
allows to construct a non-singular Wu-Yang fibration of the gauge field carried 
by the cosmic string, so that it is well defined everywhere.

How do we construct the metric that describes two accelerating bubbles? The
key is the relation that we have found between electrically charged KK
black holes
and neutral bubbles. It will be convenient to first rewrite the metric 
(\ref{eqn:boosted}) by making $r+ 2m \sinh^2\delta_Q\rightarrow r$, and
renaming
\begin{equation}
r_+\equiv 2m\cosh^2\delta_Q, \qquad r_-\equiv 2m\sinh^2\delta_Q.
\end{equation}
The metric for static electric black holes is then
\begin{equation}
\label{eqn:elecbhs}
ds^2 = \left( 1-{r_-\over r}\right)^{-1} (dx^5 +{2Q\over r} dt)^2
-\left( 1-{r_+\over r}\right) dt^2 + {1-r_-/r\over 1- r_+/r} dr^2 
+ r^2(1-{r_-\over r})^2d\Omega^2_2,
\end{equation}
with
\begin{equation}
\label{eqn:elch}
Q = {\sqrt{r_+ r_-}\over 2}.
\end{equation}
The radii $r_\pm$ must satisfy $r_+ - r_-\geq 0$, and 
$r_+ r_- \geq 0$. The mass of the black holes is
\begin{equation}
\label{eqn:elmass}
M={1\over 2} \left( r_+-{r_-\over 2} \right).
\end{equation}
The (singular) electric bubbles correspond to $0\leq -r_+ <-r_-$, and the 
neutral bubble
appears in the limiting case where $r_+=0$. In this case, the mass of the
bubble is $M= -r_-/4 >0$, and it is located at $r=0$.

In \cite{DGKT} a dilaton C-metric describing {\it magnetically} 
charged black
holes accelerating apart is given. We particularize to the KK case, and
perform a generalized
duality transformation
\begin{equation}
\label{eqn:emdual}
{\tilde F}_{\mu\nu} = {1\over 2} e^{-2\sqrt{3}\phi} 
\epsilon_{\mu\nu\rho\sigma} F^{\rho\sigma},
\end{equation}
\begin{equation}
\phi\rightarrow -\phi,
\end{equation}
while leaving the effective four-metric unchanged.
In this way we obtain the following five-dimensional metric 
describing {\it electrically}
charged accelerating black holes:
\begin{eqnarray}
\label{eqn:elecC}
ds^2 &=& {F(x)\over F(y)} (dx^5 + 2 Q y\,dt)^2 \nonumber\\ 
&+& {1\over A^2 (x-y)^2}
\left[ F(x) \left( G(y) dt^2 - {F(y)\over G(y)} dy^2 \right) 
+ F(y)^2 \left( {dx^2\over G(x)} 
+ {G(x)\over F(x)} d\varphi^2 \right) \right],
\end{eqnarray}
where $Q=\sqrt{r_+ r_-}/2$, and
\begin{eqnarray}
{}F(\xi) &=& 1+r_- A\xi, \nonumber\\
G(\xi) &=& 1-\xi^2 -r_+ A \xi^3,
\end{eqnarray}

We will see below how this metric can also describe accelerating electric
(singular) or neutral
bubbles, but it is first convenient to briefly analyze the physical
interpretation of the metric when the parameters are chosen so as to describe
black holes. 

The root of $F$ is $\xi_1=-1/(r_- A)$, and $G$ has three roots $\xi_2,
\xi_3,\xi_4$. It is
convenient to analyze them for small $r_\pm A$. Then $\xi_1$ is large and 
negative,
as is also $\xi_2\approx -1/(r_+ A)$. The other two roots
are $O(1)$: $\xi_3\approx - 1$, $\xi_4\approx +1$. 
When $r_+\geq r_-\geq 0$ their physical meaning can be 
obtained by considering the following limits:

\begin{enumerate}
\item  We blow up the region near $y=\xi_1,\xi_2$ by making $y= -1/(rA)$, 
and $A\rightarrow 0$. In this case, with the 
additional change $x=\cos\theta$ and rescaling $t\rightarrow At$, 
the metric reduces to
(\ref{eqn:elecbhs}). Alternatively, instead of saying that 
we go close to the black holes,
one could interpret this as the static limit of the metric,
where the acceleration parameter is set
to zero and the conformal factor $A^{-2}(x-y)^{-2}$
in front of the four metric is removed.

\item  While still keeping $r_\pm A\ll 1$, consider now the region near 
$y=\xi_4\approx 1$. 
This root corresponds to an acceleration horizon, since in this case,
with $F(y), F(x)\approx 1$,
the metric becomes that of Rindler space,
(though written in unusual coordinates), crossed
with a trivial internal space. Points of constant $y,x$
are moving with uniform acceleration. 
It can also be interpreted as a small particle
limit.

\end{enumerate}

This interpretation holds when $y$ lies in the range $-1/(r_-A)<y<x$. 
The endpoints of this interval are singular, and
correspond to the singularity inside the black hole and asymptotic infinity.

It is also clear that we must keep the coordinate $x$ 
between the two $O(1)$ roots of $G(x)$,
$\xi_3\leq x\leq \xi_4$. Then $x$ 
plays the role of $\cos\theta$ in the angular $(x,\varphi)$ part of the 
metric, which is topologically a sphere $S^2$. 
However, this sphere is distorted, and
can easily be seen to contain necessarily a conical
singularity at one of the poles, reflecting 
the need for an accelerating force.

{}From our discussion of the static solution above, it is clear what 
to do to obtain accelerating
bubbles. Singular electric bubbles appear for $0\leq -r_+ <-r_-$, and the 
regular neutral bubble
corresponds to $r_+=0$, $-r_- = 4M$. 
In this case
\begin{equation} 
F(\xi)=1- 4MA\xi, \qquad G(\xi) = 1-\xi^2,
\end{equation}
and all the roots are easily given: $\xi_1=-1$, $\xi_2=1$, 
$\xi_3= 1/(4MA)$. Notice that the ordering of the roots has changed. Now
we take $-1 \leq x\leq 1$ and $-\infty < y <x$. 
The axis $x=-1$ points to infinity, whereas $x=1$ points to the other 
black hole.
The value $y=-1$ corresponds
again to the acceleration horizon. 
But now, instead of a black hole horizon we find a bubble 
located at the zero of 
$F(y)^{-1}$, i.e., at $y\rightarrow -\infty$, where 
the size of the internal circles vanishes. This is just as expected, 
since by taking $y=-1/(rA)$
and $A\rightarrow 0$ we recover the neutral bubble metric, with the bubble 
at $r=0$.

We now analyze in more detail the conical singularities that pull the
bubble. The relevant part of the metric is
\begin{equation}
{}^{(2)}ds^2= {dx^2\over 1-x^2} +{1-x^2\over F(x)}d\varphi^2.
\end{equation}
For $-1\leq x\leq 1$ this is topologically $S^2$.
If the periodicity of $\varphi$ is $\Delta\varphi$, there is a
conical deficit at each of the poles $x=\pm 1$ given by
\begin{equation}
\delta_{(\pm 1)}= 2\pi\left(1 -
{\Delta\varphi\over 2\pi\sqrt{F(\pm 1)}}\right).
\end{equation}
Clearly, it can not be made zero simultaneously at both poles. 
We choose to keep regularity
along the axis $x=1$ joining the black holes, so we take
\begin{equation}
\Delta\varphi =2\pi \sqrt{F(1)}.
\end{equation}
A conical 
defect runs along the axis $x=-1$, i.e., from each black hole to infinity. 
We can 
interpret it as coming from a cosmic string with tension 
$\mu=\delta/(8\pi)$, which is pulling the black holes apart.
In the limit of small $MA$, the tension is
\begin{equation}
\label{eqn:muma}
\mu\approx MA,
\end{equation}
as expected.

We require the geometry to be regular at the bubble. 
To do so, we focus on the $(x^5, y)$ part
of the metric, in the limit $y\rightarrow -\infty$. In this case, absence 
of conical singularities requires the period of $x^5$ to be
\begin{equation}
\Delta x^5 = 16\pi M.
\end{equation}

{}Finally, we study the possibility of constructing an Euclidean instanton
describing pair production of these bubbles. 
We Wick-rotate $t\rightarrow i\tau$, and analyze what condition must be 
imposed on the Euclidean time period $\Delta \tau$ so as to avoid a conical
singularity at the acceleration horizon $y=-1$. Note that there is no
other horizon and the bubble has no prefixed temperature. 
Then it is always possible to attain regularity by setting
\begin{equation}
\Delta \tau= {4\pi \sqrt{F(-1)}\over |G'(-1)|}= 2\pi \sqrt{1+4MA}.
\end{equation}
The decay process is described as usual: take the initial configuration 
to be that of a cosmic string (i.e.,
a conical defect) in flat space. The Euclidean solution above tends 
asymptotically 
to this configuration. Then, slice in half the Euclidean instanton
at antipodal constant values of $\tau$ surfaces 
and glue it to the Lorentzian solution cut at $t=0$. This describes the 
breaking of a cosmic string with formation of neutral bubbles at its ends, 
and provides an alternative to the processes where the string breaks by
forming black holes \cite{STRINGBREAK}.

It is interesting to notice that we have been able to construct a 
C-metric instanton for a 
{\it neutral} object. This is new: usually, charge is needed to lower
the black hole horizon temperature to match the acceleration temperature. In
this case, however, we have a neutral object with zero (more properly,
non-fixed) temperature, and no such problem is present. Moreover, as far
as we know, this is the first non-extremal pair creation instanton 
constructed for a theory with dilaton coupling parameter $a>1$. In these 
theories, the black hole temperature grows without limit as one
approaches extremality, and this precludes the possibility of regularly
matching the Euclidean time periodicity. Clearly, this obstruction
disappears for bubbles.

One could also expect to be able to create neutral bubbles in a cosmological
context, in a way similar to the nucleation of Schwarzschild black holes in a
de Sitter universe \cite{GINSPARG}. However, it is very unlikely that a similar
solution for bubbles exists within the framework of pure KK theory. 
Notice that a 
cosmological constant $\Lambda$ in five dimensions would yield, upon reduction,
a four dimensional Liouville-type potential $\Lambda e^{2\phi/\sqrt{3}}$,
which is inversely proportional to the length of the internal circles. This 
potential would diverge at the position of a bubble, 
where the internal space closes up.
Such a divergence, however, would entirely come from the conformal factor in
the metric and, even if somewhat undesirable, it might give no problems in
the higher dimensional metric. We would expect to find a neutral bubble 
if we had
a KK cosmological (de Sitter) solution containing an electrically 
charged black hole. The bubble would correspond to taking the electric 
charge to be zero in the 
way described above. However, it has been argued in 
\cite{POLETTI} that no such
charged solutions exist for non-trivial Liouville-type dilaton potentials. In 
principle, this does not rule out completely the existence of neutral bubble
cosmological solutions, since they could exist and not be related to any 
electric black hole spectrum, but this certainly seems unlikely.

\subsection{Euclidean actions and pair creation rate}

As usual, by continuing the solutions to Euclidean time, their action can be 
used to obtain thermodynamical properties or pair creation rates. 
Upon dimensional reduction, the five dimensional action
\begin{equation}
\label{eqn:5Daction}
I={1\over 16\pi G_5}\int_{\widetilde{\cal M}} \sqrt{|{\tilde g}|}\; {\tilde R}
-{1\over 8\pi G_5}\int_{\partial\widetilde{\cal M}} \sqrt{|{\tilde h}|}\; 
{\tilde K},
\end{equation}
(tilded quantities will correspond to the 5D description) turns to
\begin{equation}
\label{eqn:4Daction}
I={1\over 16\pi G_4}\int_{{\cal M}} \sqrt{|g|}[R - 2(\nabla \phi)^2 
-e^{-2\sqrt{3}\phi} F^2]
-{1\over 8\pi G_4}\int_{\partial{\cal M}} \sqrt{|h|}\; {K}.
\end{equation}
The boundary $\partial{\cal M}$ is taken to be the one induced by quotienting 
the five dimensional boundary $\partial\widetilde{\cal M}$.
Also, the five dimensional and four dimensional coupling constants are related
by $G_4= G_5/\Delta x^5$.
{}For exact solutions of the Kaluza-Klein equations, the volume term in
(\ref{eqn:5Daction}) vanishes and only the boundary term remains. Dimensional
reduction of this term yields
\begin{eqnarray}
\sqrt{|{\tilde h}|}\; {\tilde K} &=& \sqrt{|h|}\; 
{e^{-\phi/\sqrt{3}}\over\sqrt{|g|}}
\partial_\mu(e^{\phi/\sqrt{3}}\sqrt{|g|}\; n^\mu)\nonumber\\
&=& \sqrt{|h|}\; K + \sqrt{|h|}\; n^{\mu}\partial_\mu
\ln(e^{\phi/\sqrt{3}}\sqrt{g}),
\end{eqnarray}
where $n$ is the normal vector to the boundary $\partial{\cal M}$ induced
by the normal ${\tilde n}$ to $\partial\widetilde{\cal M}$,
\begin{equation}
{\tilde n}\rightarrow e^{-\phi/\sqrt{3}} n.
\end{equation}
Thus we see that the result not only contains
the extrinsic curvature boundary term in (\ref{eqn:4Daction}), but also a
surface term involving the scalar field. The latter can be obtained from 
(\ref{eqn:4Daction}) on-shell by substituting the scalar field equation of
motion, as is done in \cite{DGGH}.

We have remarked that the bubbles are, in the four dimensional effective
description, naked singular points. However, when computing the action
one should not
include an additional boundary surrounding the singularity, since this
must not be introduced in five dimensions, where the geometry is regular. As 
we have just explained, once the boundary terms at infinity have
been properly 
accounted for, the results should be the same either if 
we work in 5D or 4D.

{}From the Euclidean continuation of the static metrics 
in Sec.~\ref{sec:KKmnpls} we obtain their thermodynamical properties, 
such as the entropy.
The static
KK Schwarzschild solution and the neutral bubble correspond to, 
in the Euclidean regime, 
the same metric. However, the Schwarzschild black hole has the usual 
Bekenstein-Hawking entropy, whereas no entropy is associated to the bubbles. 
Of course, 
there is nothing paradoxical here, since the entropy depends on the choice of
the periodic variable (Euclidean time) that is associated to the 
asymptotic temperature.
{}For both the Euclidean continuation $t\rightarrow i\tau$
of the bubble solution (\ref{eqn:hole})
and Euclidean Schwarzschild$\times (dx^5)^2$, 
the action is\footnote{From now on we set $G_4=1$. Thus
$\Delta x^5$ does not appear in (\ref{eqn:stact}).}  
\begin{equation}
\label{eqn:stact}
I={1\over 2} m\Delta\tau.
\end{equation}
However, for the bubble, $M_B=m/2$, whereas for Schwarzschild $M_S=m$. Then,
setting $\Delta\tau=\beta$,
\begin{equation}
I_{\rm Bub}= \beta  M_B,\qquad I_{\rm Sch}={1\over 2}\beta  M_S.
\end{equation}
The vector $\partial_\tau$ has a fixed point at the horizon for the 
Schwarzschild solution, but it acts freely on the bubble. 
In the latter case, $\beta$ and $M$ can be varied independently
while remaining on-shell. The usual reasoning then leads to zero entropy 
for the
bubble, and Bekenstein-Hawking entropy $S=A/4$ for the black hole.
As we have stressed, for both solutions the only boundary 
is the surface at infinity.

The action of the analytically continued C-metric gives the dominant
contribution $e^{-I}$ to the semiclassical bubble pair 
production by snapping strings. The calculation requires 
carefully matching the boundaries near infinity $x=y=\xi_1=-1$
of the C-metric and the reference
background (flat space with a cosmic string), see e.g. \cite{HHR}.
Then, straightforward application of the formulae above for the action gives 
the result
\begin{equation}
\label{eqn:caction}
I={\pi\over 4 A^2} |F'(-1)|\sqrt{F(1) F(-1)}={\pi M\over A}\sqrt{1- 16M^2 A^2}.
\end{equation}
This answer can also be easily obtained by decomposing the
action as 
\begin{equation}
I=\beta H-{1\over 4}\Delta A,
\end{equation}
where $H$ is the physical Hamiltonian and $\Delta A$ is the difference in
the areas of the acceleration horizons of the C-metric and the reference 
background \cite{HHR}. 
As it turns, $H=0$, and the difference in acceleration horizon
areas is found to account for the entire action (\ref{eqn:caction}). 
Again, no
boundary conditions have to be imposed at the 4-dimensional singularity.

The exact relation between $A$ and $\mu$ can be inverted to express the 
action in terms of only $M$ and $\mu$:
\begin{equation}
I={\pi M^2\over\mu}{1-4\mu\over 1-2\mu}.
\end{equation}
{}For small $\mu$, the leading term is the same as for string breaking 
with monopole or black hole formation.
Subleading corrections are due to graviton and scalar exchange between
different points of the Euclidean circular trajectory of the bubble.

\section{Pair creation of magnetic bubbles and Dyson's instability}

Now that we know how to pair create neutral bubbles, it would be natural to try
and look for pair creation of magnetic bubbles. 
Such process could be readily described if we had a KK C-metric for
{\it dyonic} black holes. Our analysis in Sec.~\ref{sec:KKmnpls} 
shows that
bubbles are easily obtained from black hole solutions
by sending the electric charge to zero in
the appropriate way, i.e., by taking an infinite velocity boost 
in the internal direction. 
As we have just seen in the previous section, this
method certainly works for C-type metrics. Thus, we believe that a 
C-metric for magnetic bubbles can be constructed.

However, the Kaluza-Klein dyonic C-metric is likely to be
very complicated, since it must combine the intricacies of both the
static dyonic solutions (to which it must reduce in the limit 
$A\rightarrow 0$),
and of the C-metrics. The full dyonic C-metric solution must simplify greatly
when particularized to describe accelerating magnetic bubbles, 
so one could try to find first the metric for the latter, which,
by the way, is the situation of interest to us now. Even with this 
simplification, the solution is probably not as simple as any of the
known C-metrics, and we have not
found it yet. Some steps towards its
construction are described in Appendix \ref{apb}.

Although we do not have the explicit form of the solution, we can 
semiquantitatively describe
several features it must certainly have. Just like in the neutral case, this 
C-metric must
contain conical singularities pulling apart the bubbles. Since the
bubbles are magnetically charged, these singularities
can be removed by introducing a background magnetic field $B$ asymptoting to
the Kaluza-Klein Melvin solution, which then provides the required force. 
In Kaluza-Klein theory there is a simple way to do
this: when identifying points in the internal direction, 
introduce a twist
along the axial rotation axis \cite{DGKT}. The solution thus
obtained could be used to describe a new decay mode of magnetic fields 
in Kaluza-Klein theory. To 
leading order, the action of the instanton should be given by the Schwinger
value
\begin{equation}
I={\pi M^2\over PB}+\cdots
\end{equation}
and would be smallest for the minimal mass bubble, i.e., the extremal 
GPS monopole. Also,
the entropy enhancement factors of non-extremal black holes should be absent.

Having a metric that describes
two accelerating magnetic bubbles, then, as we have seen, we could easily
obtain from it a metric corresponding to two accelerating two-timing monopoles. 
In particular, the massless case is of great interest, since 
no force is needed to accelerate the monopoles, and therefore, no conical 
singularities should appear. That this no-external-force condition 
can be actually met has been explicitly verified for accelerating 
massless black holes in \cite{RASHEED,RE2}.
The anti-Maxwell theories considered in \cite{RASHEED} include
the two-timing KK theory for
a specific value of the dilaton.
However, the attempt at constructing a regular Euclidean instanton 
describing Dyson's
pair creation of anti-Maxwell massless black holes failed. There are no
Einstein-antiMaxwell-dilaton extremal black holes, and it 
is not possible to lower down the 
black hole horizon temperature to a value small enough to match the 
acceleration temperature \footnote{It has been pointed out
by S.~Hawking that this absence of an extremal limit is not
as odd as it may appear: in anti-Maxwell
theory, the photons emitted by a black hole have negative energy, and 
therefore a black hole should increase its mass by radiating them out!
\cite{SWH}}. Remarkably enough, this obstruction should be
absent for the monopole bubbles, since they have no horizon
and no pre-fixed temperature. Thus, the two-timing vacuum would not
decay by production of massless black holes, but rather by forming zero
mass monopole pairs.

{}For a general massive two-timing monopole, if no background 
field is introduced and, therefore, the monopoles are pulled by strings, 
we expect
the action of the instanton to be given by the usual effective low
energy result
\begin{equation}
I={\pi M^2\over \mu} [1+O(M)] = {\pi M\over A} +O(M^2).
\end{equation}
The massless monopole configuration is smoothly connected to the positive 
mass solutions, and one could perhaps expect the action to vanish in the 
zero mass limit. However, it is not at all clear that this should
be so. In fact, recently
a Euclidean C-metric instanton with massless black holes has been 
found to have {\it negative action} \cite{RE2}! 
If we recall that two
oppositely charged two-timing monopoles repel each other, an enhanced 
(instead of suppressed) pair creation rate would not come as big surprise.
We can see that, in any case,
pair creation of massless two-timing
monopoles would imply a crass instability of the 
two-timing vacuum. 

On the other hand, we could also consider a solution with accelerating 
negative mass monopoles. This solution can asymptote to the two-timing
flat vacuum if we allow for a conical deficit inbetween the monopoles: the
singularity would pull them together, but, since they have negative mass, 
they accelerate in the opposite direction. The conical deficit
can operationally be replaced by a cosmic string like in \cite{RE}, 
and in this way we would
find a decay of the two timing vacuum by spontaneously producing 
cosmic strings with negative mass monopoles at their ends. 

Notice that in 
two-timing decay processes the tunneling takes place through a path that,
though effectively Euclidean in four dimensions, is Lorentzian
in the full five dimensional picture. 
In this case it is not completely clear how the classical 
action contributes to the leading terms of the decay rate. 
A possible resolution is to simply take the point of view of the effective four
dimensional description and write $\Gamma\sim e^{-I}$. 

As usual, assuming a supersymmetric vacuum  will rule out the decay 
of the vacuum via the
new C-metric instanton which we have attempted to find here.  
More explicitly, the
instanton will be simply connected and so the only spin structure 
allowed will be the one for
which the fermions satisfy aperiodic boundary conditions as 
they are parallel propagated
around the internal space.  However, for supersymmetry we 
would require the fermions
to be periodic on the internal space, and therefore it would not be possible 
to match the fermion structures when trying to construct the pair
creation instanton.
As in \cite{WITTEN}, this is the most 
natural mechanism for 
suppressing this pathological behaviour.

\section{Discussion and conclusions}

The bubble solutions of the form $(-dt^2)\times$Euclidean-Taub-NUT in 
Sec.~\ref{sec:KKmnpls} were already constructed some time ago. However, 
we have found in our analysis that they have a common origin with black 
holes, and can be constructed out of the dyonic black hole solutions. 
This has allowed us
to extend the class of pair creation processes studied till now. 
Interestingly, we have found that non-extremal
pair creation instantons are indeed possible 
in a theory with dilaton coupling $a>1$.

The (one-timing) bubbles studied here suffer 
from classical instabilities, but so do many extended black objects. 
The bubbles of Sec.~\ref{sec:KKmnpls} had been constructed in previous
literature \cite{SORKIN,GROSS}
by adding a time coordinate to formerly known gravitational instantons. 
However, our approach to constructing them by identifying the solutions
in the parameter space of black hole
solutions is much more general, and can be applied to other theories
containing internal dimensions. 
In this way one can find
bubbles--including also bubble strings--as solutions of the low
energy effective equations of string theory, and of other higher dimensional 
theories. These will be discussed elsewhere. 

Given the finite size of bubbles, it is natural to ask whether rotating
bubbles exist. However, this does not seem likely to occur, at least in 5D KK. 
If we consider the rotating dyonic solutions of this theory \cite{ROTDYON},
then any way of eliminating the electric charge to obtain bubbles (such as
taking boosts $v\rightarrow \infty$) also sends the angular momentum of the
solution to zero.
In fact, the bubble solutions thus obtained take the form of 
$(-dt^2)\times$Euclidean-Kerr-Taub-NUT. These do not correspond to 
rotating bubbles, but, rather, to objects in background magnetic fields 
\cite{DOWKER}. When the
``Euclidean-rotation'' parameter $\Omega$ is small as compared to the inverse
compactification scale, the four dimensional description is that of a 
distorted bubble in a background magnetic field. The bubble has a magnetic 
dipole moment, in addition to the monopolar charge it may already have. 
On the other hand, if $\Omega$ is comparable to the inverse compact radius, 
then the appropriate
picture is in terms of two GPS monopoles at antipodal points 
of the Euclidean $S^2$ bolt, with, in general, charges 
of different magnitude. In all these configurations, conical singularities 
appear so as to keep the
whole set static. 

A peculiar outcome of our analysis is the massless monopole in the two-timing
theory. It is likely that massless objects are a 
rather general feature of theories
with internal times. Usually, the mass and charges are expressable as sums and
products, respectively, of several characteristic radii, like in 
(\ref{eqn:elch}-\ref{eqn:elmass}). 
The imaginary charge associated with an internal time 
may be obtained by changing
the sign of one of these radii, and this, in turn, may lead to regions of
parameter space where the mass can be set to zero.

On the other hand, classical massless black hole solutions 
have been recently found in 
the context of string theory \cite{MASSLESS}. 
They exhibit a number of peculiar features,
most of them associated to the fact that they are nakedly 
singular \cite{RE2}.
Our massless monopoles exhibit also a singularity in four dimensions, but 
this is harmless since we know how to handle it by going to 
the regular five dimensional manifold.
Another
significant difference is that, whereas the stringy massless holes are
supersymmetric, the two-timing massless monopoles are 
not. 
As a consequence, they are expected
to acquire mass by quantum corrections. Given the presence of negative
mass states, it is not clear what the quantum
corrected mass spectrum of monopoles would be. As we have argued, the
massless monopole could be stabilized due to the symmetry between the two
asymptotic regions it connects. However, the quantization of
the two-timing theory is certainly bound to exhibit all kinds of
pathologies. 

{}Finally, we have presented evidence that theories with 
internal time directions
can give rise to pair production of massless particles out of the
vacuum. This evidence is
based on the following facts: (i) The 5D KK theory contains a 
non-singular massless monopole,
which can be obtained by appropriately choosing parameters in the general
dyonic solution; (ii) Dilaton C-metrics for both electric and magnetic black 
holes are known, so we expect that a similar metric 
for dyonic black holes must exist;
(iii) We have explicitly shown that it is possible to construct, 
using the C-metrics already known, instantons for
pair production of bubbles.

Thus, theories with extra times are expected to suffer from
an instability that in some sense is even worse than those previously known. 
Pathologies caused by negative norm
fluctuations of the internal dimension can be kept below a certain level by, 
for example, 
taking the internal 
radius to be sufficiently small. On the other hand, 
the instability of the ordinary KK vacuum described in 
\cite{WITTEN} is suppressed for large internal 
radius\footnote{Notice that this decay mode 
cannot be generalized to the two-timing theory, since by Wick-rotating 
the internal direction of the instanton we find 
the 5D {\it Lorentzian} Schwarzschild 
solution, which is {\it not} regular.}. In contrast,
production of massless two-timing monopoles could be essentially unsuppressed
(or even enhanced, as in \cite{RE2}). 
Work towards the explicit construction of the corresponding C-metric, 
as well as the more general dilaton dyonic C-metric, is in progress.

\acknowledgements

A.C. was supported 
by NSF PHY94-07194. R.E. was partially supported by a FPI 
postdoctoral fellowship from MEC (Spain),
and by CYCIT AEN-93-1435 and UPV 063.310-EB225/95.

\appendix

\section{}\label{apa}

{}For some purposes, it is useful to write the solution (\ref{eqn:gensol}), 
(\ref{eqn:XYf}) as it is done in \cite{WILTSHIRE}, 
in terms of the more physical parameters $M,P,Q$, and the scalar
charge $\Sigma$. The latter is obtained as $\phi\rightarrow \Sigma/r + O(r^{-2})$
and is 
not an independent parameter, but rather it must satisfy
\begin{equation}
\label{eqn:scalar}
{2\over 3}\Sigma = {Q^2\over \Sigma-\sqrt{3}M}+ {P^2\over \Sigma+\sqrt{3}M}.
\end{equation}
The functions (\ref{eqn:XYf}), after a linear shift of $r$, are given by
\begin{eqnarray}
X &=& (r-\Sigma/\sqrt{3})^2+{2Q^2\Sigma \over \Sigma-\sqrt{3}M},\nonumber\\
Y &=& (r+\Sigma/\sqrt{3})^2+{2P^2\Sigma \over \Sigma+\sqrt{3}M},\\
f &=& (r-M)^2 -(M^2 + \Sigma^2 - Q^2 -P^2),\nonumber
\end{eqnarray}
and 
\begin{eqnarray}
A_t &=& Q{r+\Sigma/\sqrt{3}\over X},\nonumber\\
A_\varphi &=& P\cos\theta.
\end{eqnarray} 

This representation has the advantage that the electric-magnetic duality 
(\ref{eqn:emdual}) of the four dimensional description is
evident from the symmetry under $Q\leftrightarrow P$,
$\Sigma\rightarrow -\Sigma$. 

At first sight, it is not clear how to obtain
the bubble sector of the solution, since if we directly set $Q=0$ we
only find the magnetically charged black holes. However, a different solution
is obtained if we send $Q\rightarrow 0$ 
and simultaneously $\Sigma\rightarrow \sqrt{3}M$, while keeping the ratio 
$Q^2/(\Sigma- \sqrt{3}M)$ finite
and fixed by (\ref{eqn:scalar}).
In this limit we find
\begin{eqnarray}
X &=& f = (r-M-\sqrt{4M^2-P^2})(r-M+\sqrt{4M^2-P^2}), \nonumber\\
Y &=& (r+M-P)(r+M+P).
\end{eqnarray}
Shifting now $r+M\rightarrow r$ we find the same metric as in (\ref{eqn:5DTN}).

\section{}\label{apb}

In this appendix we find a unified form for the several known limits that
the C-metric for magnetic bubbles must satisfy. 
The full solution has not been found yet, 
but we think that the results presented here may be useful
for further investigations.

The KK C-metric for magnetic bubbles will depend on three parameters, 
namely $r_+,r_-, A$. It must certainly satisfy the following limits: 

\begin{enumerate}
\item  For vanishing $r_+$ or $r_-$ it must reduce to the neutral 
bubble C-metric.

\item  For $r_+=r_-$ it must reproduce the C-metric describing creation of 
GPS monopoles \cite{DGGH}.

\item For $A\rightarrow 0$ it must reduce to the static general
Taub-NUT solution cross a trivial time. 

\end{enumerate}

Of course, these metrics need not be reproduced in precisely the form given
above or in \cite{DGGH}, and some coordinate changes may be needed.
We will rewrite all these metrics in a way that a unified
form all of them is possible. 

Start first with the static limit, i.e., Taub-NUT$\times (-dt^2)$. 
Writing it as
\begin{eqnarray}
\label{eqn:statlim}
ds^2 &=& -dt^2 +{(r-r_+)(r-r_-) \over r^2-r_+r_-}[dx^5 +2\sqrt{r_+ r_-}
\cos\theta d\varphi]^2 \nonumber\\
&+& {r^2-r_+r_-\over(r-r_+)(r-r_-)}dr^2 + 
(r^2-r_+ r_-)d\Omega_2^2,
\end{eqnarray}
we change to the coordinate 
\begin{equation}
y=-{1\over (r+r_+)A}
\end{equation}
(other choices are possible, like $y=-1/(rA)$ or $y=-1/((r\pm r_-)A)$, but this
one is quite convenient since it will allow for the simplest form of the
off-diagonal term in the C-metrics with magnetic charge). 
In this static metric 
the parameter $A$ has no special 
meaning, and simply has the dimensions of an inverse length.

Then, the metric (\ref{eqn:statlim}) is
\begin{equation}
ds^2 = {H(y)E(y)\over J(y)} (dx^5 +2\sqrt{r_+ r_-}
\cos\theta d\varphi)^2 -dt^2 + {J(y)\over H(y) E(y)}{dy^2 \over A^2 y^4} + 
{J(y)\over A^2 y^2} d\Omega_2^2,
\end{equation}
where we have defined the following polynomial functions
\begin{eqnarray}
H(\xi) &=& 1+2r_+ A\xi,\nonumber\\
E(\xi) &=& 1+(r_+ +r_-) A\xi,\\
J(\xi) &=& 1+2r_+ A \xi + r_+(r_+-r_-) A^2\xi^2.\nonumber
\end{eqnarray}
The following cases are specially interesting:

\begin{enumerate}

\item For $r_+=r_-$ we have $H(y)=E(y)=J(y)=1+2r_+ A y$. The nut is at 
$y=-1/(2r_+ A)$.

\item For $r_+=0$, we have $H(y)=J(y)=1$, $E(y)=1+r_- Ay$. The bubble
is at $y=-1/(r_-A)$.

\item For $r_-=0$, $E(y)=1+r_+ Ay$, $J(y)=E(y)^2$. The bubble is at
$y=-1/(2r_+ A)$.

\end{enumerate}

We can now rewrite the KK C-metrics with bubbles or GPS monopoles in a way
that the bubbles or monopoles are located at precisely these same values 
of $y$. This 
requires that in the C-metrics we make coordinate changes of the form
\begin{eqnarray}
\label{eqn:change}
y^{-1}\rightarrow y^{-1} + r_0 A, \nonumber \\
x^{-1}\rightarrow x^{-1} + r_0 A,
\end{eqnarray}
where by adequately choosing $r_0$ we can locate the bubbles where required.

After some straightforward manipulations, 
we can write all these metrics in the
following unified form:
\begin{eqnarray}
\label{eqn:unifiedC}
ds^2 &=& {H(y)E(y)J(x)\over H(x)E(x)J(y)} (dx^5 +2 x \sqrt{r_+ r_-}
d\varphi)^2 \nonumber\\
&+& {1\over A^2(x-y)^2} \biggl[ H(x)E(x) \left( K(y) dt^2 - 
{J(y) \over E(y)H(y) K(y)} dy^2 \right) \nonumber\\
&+& J(y)\left( {dx^2\over K(x)} + {K(x)E(x)H(x)\over J(x)} d\varphi^2 
\right) \biggr].
\end{eqnarray}
The functions $H,E,J$ are as given above. Despite its complicated aspect,
this metric easily reproduces all the required limiting solutions if the 
function $K$ is properly chosen. Precisely, one must take

\begin{enumerate}

\item For $r_+=r_-$:
\begin{equation}
K(\xi) = {1\over H(\xi)} -\xi^2.
\end{equation}

\item For $r_+=0$:
\begin{equation}
K(\xi) = E(\xi)^2 -\xi^2.
\end{equation}

\item For $r_-=0$:
\begin{equation}
K(\xi) =H(\xi)^2 -\xi^2.
\end{equation}

\end{enumerate}

Whatever the C-metric for magnetic bubbles is, it must be possible to write
it in a way that these solutions are found.
Additionally, the static limit $A\rightarrow 0$ requires that, for large $y$, 
$K(y)\sim -y^2$, and for $x=O(1)$, $K(x)\sim 1-x^2$. Notice that there 
are two forms of a 
neutral bubble C-metric; they are related by a coordinate change 
like in (\ref{eqn:change}). Alternative forms are possible, but the magnetic
potential term gets more complicated.
Unfortunately, we have not found any simple enough function that reproduces the 
above limits and, 
simultaneously, solves the five-dimensional Einstein's equations.

If we tried to aim at
finding only the most interesting case, 
the solution containing massless monopoles,
the approach just described might not be adequate, and
other simplifications may be more useful. For massless objects,
the polynomial functions in the C-metric have to be even (or, more
properly,
one expects them to be even for at least some choice of coordinates). 
This property
can easily be seen to be related to the no-external-force condition in
the $(x,\varphi)$ part of the metric, or equivalently, the vanishing 
of the mass
in the (static limit of the) $(t,y)$ part, see \cite{RE2}. 
However, even in this particular case the
solution has remained elusive.

\end{document}